\documentclass[12pt]{iopart}
\usepackage{graphicx}
\usepackage{amssymb}
\begin{document}

\title[Growth and post-annealing studies of La-Bi2201 single crystals]
{Growth and post-annealing studies of
Bi$_2$Sr$_{2-x}$La$_x$CuO$_{6+\delta}$ (0$\leq x \leq$1.00) single
crystals}

\author{Huiqian Luo, Peng Cheng, Lei Fang, Hai-Hu Wen}

\address{
National Laboratory for Superconductivity, Institute of Physics and\\
National Laboratory for Condensed Matter Physics, P. O. Box 603
Beijing, 100190, P. R. China }

\ead{hhwen@aphy.iphy.ac.cn}

\begin{abstract}
Bi$_2$Sr$_{2-x}$La$_x$CuO$_{6+\delta}$ (0$\leq x \leq$1.00) single
crystals with high-quality have been grown successfully using the
travelling-solvent floating-zone technique. The patterns of X-ray
diffraction suggest high crystalline quality of the samples. After
post-annealing in flowing oxygen at 600 $^o$C, the crystals show
sharp superconducting transitions revealed by AC susceptibility. The
hole concentration $p$ is deduced from superconducting transition
temperature ($T_c$), which exhibits a linear relation with La doping
level $x$.  It ranges from the heavily overdoped regime ($p \approx$
0.2) to the extremely underdoped side ($p \approx$ 0.08) where the
superconductivity is absent. Comparing with the superconducting dome
in Bi$_{2+x}$Sr$_{2-x}$CuO$_{6+\delta}$ system, the effects from
out-of-plane disorders show up in our samples. Besides the La doping
level $x$, the superconductivity is also sensitive to the content of
oxygen which could be tuned by post-annealing method over the whole
doping range. The post-annealing effects on $T_c$ and $p$ for each
La doping level are studied, which give some insights on the
different nature between overdoped and underdoped regime.
\end{abstract}

\section{Introduction}
The high temperature superconductivity (HTSC) in copper oxides is
one of the most important issues in condensed matter physics.  In
hole doped cuprates, the antiferromagnetic Mott insulator at half
filling will be gradually replaced by a novel metallic state when
the hole concentration $p$ increases \cite{Lee}. The hole doping
dependence of superconducting transition temperature ($T_c$) forms a
superconducting dome in the phase diagram \cite{Preslan}, which is
one of the central challenges to explain the mechanism of HTSC.

The single-layered cuprate Bi$_2$Sr$_2$CuO$_{6+\delta}$ (Bi2201) is
one of the well studied materials for the mechanism of HTSC due to
its advantage of relative lower $T_c$. The coordination of Sr$^{2+}$
ions located next to the apical oxygen, which is defined as the
A-site, could be occupied by the ions of lanthanide (Ln$^{3+}$) and
bismuth (Bi$^{3+}$). It generates a large family including
Bi$_2$Sr$_{2-x}$Ln$_x$CuO$_{6+\delta}$ (Ln= La, Pr, Nd, Sm, Eu, Gd,
etc.) systems and Bi$_{2+x}$Sr$_{2-x}$CuO$_{6+\delta}$ (Bi-Bi2201)
system \cite{Eisaki}. The mismatch between the size of substituted
ions introduces A-site disorders, which result in different $T_c$
ranging from 30 K to 10 K at the optimal doping level in each system
\cite{Fujita}. The Bi$_2$Sr$_{2-x}$La$_x$CuO$_{6+\delta}$
(La-Bi2201) system has the weakest influence of disorders and the
highest $T_c$ in this family, while the case in the
Bi$_{2+x}$Sr$_{2-x}$CuO$_{6+\delta}$ system is in the opposite way.
The relationship between the disorders and $T_c$ gives more
opportunities to access the central problems in the mechanism of
HTSC \cite{Okada,Slezak}, and single crystals with high quality are
highly desired for such investigations.

It was reported that the Bi2201 single crystals could be grown by
the self-flux and KCl-solution-melt method
\cite{Michel,Fleming,Martovitsky,Gorina}. However, as far as we
know, sizeable crystals with high homogeneity and less contamination
could be obtained only through the optical travelling-solvent
floating-zone (TSFZ) method \cite{Liang1,Liang2,Ando1}. We have
successfully grown high-quality Bi$_{2+x}$Sr$_{2-x}$CuO$_{6+\delta}$
single crystals with highly dense doping levels in the underdoped
regime.  The characterizations show that our crystals have high
crystalline quality and sharp superconducting transitions.  The
quick contraction of $c$-axis accompanied with the suppression of
superconductivity suggests that more disorders are introduced by
doping Bi$^{3+}$ into the Sr-O plane \cite{Luo1}.

Besides the ionic substitution in the out-of-plane, it is known that
the hole concentration of cuprates also controlled by the oxygen
content which could be tuned by post-annealing method. When a sample
is annealed at a certain temperature and atmosphere, the activated
oxygen atoms redistribute inside the sample. The holes on O sites in
Cu-O plane will be added or removed through these processes, thus
the superconductivity could be tuned sensitively. However, there are
rare reports on the post-annealing effects over the whole doping
range in La-Bi2201 system, and the detailed behaviors for
post-annealing are not clear yet \cite{Lin,Ando2}.

In this paper, we report the successful growth of
Bi$_2$Sr$_{2-x}$La$_x$CuO$_{6+\delta}$ single crystals with high
quality by TSFZ method.  The nominal composition  of La doping level
$x$ varies from 0 to 1.00,
 and the hole concentration of our
crystals ranges from the heavily overdoped regime to the extreme
underdoped side where the superconductivity is absent.  The
post-annealing experiments have been extensively carried out on the
samples at each La doping level at different temperatures in flowing
oxygen. The post-annealing effects on $p$ and $T_c$ are studied,
which give different insights between underdoped and overdoped
regime.

\section{Experiments}

The Bi$_2$Sr$_{2-x}$La$_x$CuO$_{6+\delta}$ (0$\leq x \leq$1.00)
single crystals were grown by the TSFZ technique.  The starting
materials were prepared by a standard solid state reaction method
before the crystal growth.  In order to eliminate the moisture, the
powders of SrCO$_3$ (99.99\%) were baked at 200 $^o$C for more than
5 hours. And the powders of La$_{2}$O$_3$ (99.99 \%) were calcined
at 1000 $^o$C for 10 hours to make sure that La(OH)$_3$ has
decomposed completely.  After the pre-processes, they were mixed
with Bi$_{2}$O$_3$ (99.99 \%) and CuO (99.5\%) in the stoichiometric
proportion (2-x) : x/2 : 1 : 1. The mixed powders were ground by
hand in a dry agate mortar for about 4 hours, and then they were
calcined in a crucible at 800 $^{o}$C for 24 hours inside a muffle
furnace. As soon as the product was cooled down to room temperature,
it was crushed into powder then ground and calcined again. This
procedure was repeated for four times to ensure the homogeneity of
the starting material. Finally, the homogenous polycrystalline
powder was pressed into a cylindrical rod of $\phi 7$ mm $\times$ 90
mm under hydrostatical pressure at $\thicksim$70 MPa, then it was
sintered in a vertical furnace at 850 $^{o}$C for 36 hours in air.
In order to get a feed rod with higher density and sufficient oxygen
content, the premelting process was performed under oxygen pressure
$P(O_2)=2$ atm, and the moving speed of mirror stage was 30 mm/hr.
After premelting, a homogeneous feed rod with $\phi$6$\sim$7 mm in
diameter and 60$\sim$80 mm in length was obtained.

Single crystal growth by the TSFZ method was performed with an
optical floating-zone furnace which was produced by the
\emph{Crystal Systems Corporation}. A steep temperature gradient was
obtained by using four 300 W halogen lamps as the heating sources.
The crystal growth was done under an oxygen pressure with 2.0 $\sim$
3.5 atm in an enclosed quartz tube, and the flowing rate of O$_2$
was about 20 $\sim$ 40 cc/min. The typical growth rate is about 0.50
mm/hr, and the rotation rate is 25.0 rpm for the upper shaft and
15.0 rpm for the lower shaft in opposite directions. The
post-annealing was carried out under 1 atm flowing oxygen at a
finite temperature for more than 150 hours at each step. At the end
of treatment, the samples were quenched to room temperature in air.

The crystals cleaved from the as-grown ingots were selected
carefully under polarization microscope, and then characterized by
various techniques. The X-ray diffraction (XRD) measurements of the
crystals were carried out by a \emph{Mac- Science} MXP18A-HF
equipment with $\theta-2\theta$ scan to examine the crystalline
quality of the samples. $K_{\alpha}$ radiation of Cu target was
used, and the continuous scanning range of 2$\theta$ is from 5$^{o}$
to 80$^{o}$. The composition of crystal was examined by the energy
dispersive X-ray (EDX, Oxford-6566) analysis. The superconductivity
of the crystals was characterized by AC susceptibility  based on an
\emph{Oxford }cryogenic system Maglab-EXA-12 and a \emph{Quantum
Design} Magnetic Property Measurement System (MPMS).  An alternating
magnetic field $H=0.1$ Oe was applied perpendicular to the
$ab$-plane with a frequency $f$=333 Hz during the AC susceptibility
measurement.  The transition temperature of the samples was derived
from AC susceptibility curve by the point where the real part of the
susceptibility deviates from the flattened normal state part.

\section{Results and discussion}

\subsection{Crystal growth and characterization}
\begin{figure}
  \center\includegraphics[width=2.8in]{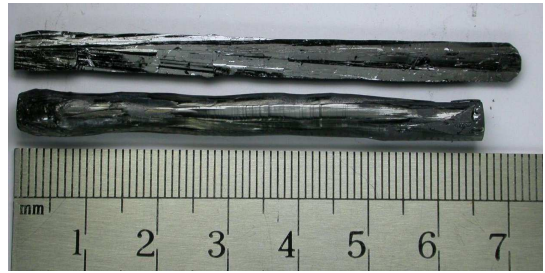}\\
  \caption{(color online) One of the as-grown ingots with
$x$= 0.85 (below) and the cleaved ingot with $x$= 0.40 (above).}
  \label{f1}
\end{figure}
The crystal growth was performed under oxygen atmosphere enclosed by
a quartz tube.  In order to obtain high quality single crystals, the
shape of melting-zone should be rather stable during the growth.
Thus the power of lamp, growth rate, rotation rate of shafts,
pressure and flowing rate of oxygen should be carefully tuned.
According to our experiences, the melting point of feed rod
increases when more and more La$_{2}$O$_3$ is mixed into the staring
material, and it is also related to the oxygen pressure.  Two kinds
of pressure were applied: $P(O_2$)= 2.2 atm for underdoped samples
with 0.40 $\leq x \leq$ 1.00 and $P(O_2$)= 3.4 atm for overdoped
samples with 0$ \leq x \leq$ 0.30, respectively. The other
parameters are almost the same for each nominal composition during
the growth. Large and plate-like single crystals could be obtained
easily for $x \geq$ 0.70. When $x$ reduces to 0.40 $\sim$ 0.60, the
cleaved as-grown crystals are lamellar.  For the overdoped samples
with $x <$ 0.40, the growth of crystals are much more difficult.
Since the viscosity of the melted compound is low during the growth,
only needle-like crystals could be obtained for the compositions
with $x$= 0 and 0.05. One of the as-grown ingots with $x$= 0.85
(below) and the cleaved ingot with $x$=0.40 (above) are shown in
Figure 1. The cleaved crystals from them are sizeable and flat in
large area.

\begin{figure}
   \center\includegraphics[width=3.0in]{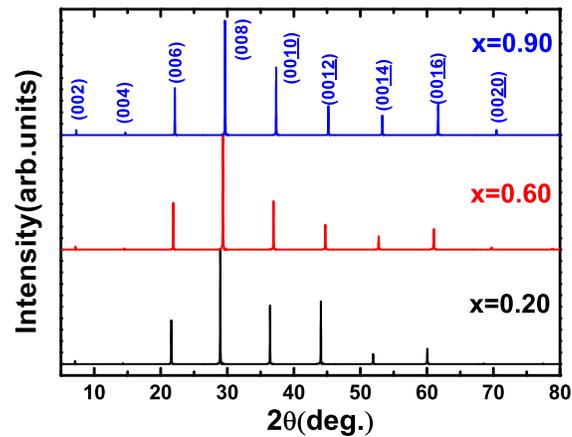}\\
  \caption{(color online) Typical XRD patterns for cleaved single crystals with $x$= 0.20, 0.60, 0.90. All peaks are very sharp
with narrow FWHM around 0.1$^{\circ}$.}
  \label{f2}
\end{figure}
The crystal structure was examined by XRD measurement with incident
ray along the $c$-axis of the single crystal.  Some typical
diffraction patterns are shown in Figure 2.  All peaks are along
(00$l$) with a narrow full-width-at-half-maximum (FWHM) around
0.1$^{\circ}$, which shows high crystalline quality and $c$-axis
orientation in our samples. The $c$ lattice parameters deduced from
these patterns are about 24.61 \AA, 24.24 \AA\ and 24.03 \AA\ for
$x$= 0.20, 0.60 and 0.90, respectively. The small variation of
$c$-axis for different La doping level reveals the ionic
substitution effects in the out-of-plane.

\subsection{Hole concentration and Superconductivity}

The superconductivity of the crystals was characterized by the
measurements of AC susceptibility.  The La-Bi2201 system is so
sensitive to a low magnetic field that the superconducting
transition could be suppressed by a tiny DC field as low as 1 G.
Figure 3(b) shows that the superconductivity is suppressed quickly
under low external DC fields. In order to obtain the exact magnitude
of $T_c$ under zero external DC field, the residual field in the
external superconducting coils should eliminated completely before
the measurements. In our experiments, the external persistent DC
field is lower than 0.5 G, and the results are more reliable and
repeatable. Because of the high anisotropy and inhomogeneity in
Bi2201 systems and segregation during the growth, the
superconducting transition is broad for the as-grown crystals. The
$T_c$ also has a broad distribution among different crystals cleaved
from the same ingot.  However, after post-annealing in flowing
oxygen for more than 150 hours, the superconducting transition gets
much sharper, and the $T_c$ among the samples with the same nominal
composition is almost the same. Figure 3(a) shows the typical AC
susceptibility data for the samples annealed at 600 $^o$C with
$x$=0.05, 0.10, 0.20, 0.40, 0.60 and 0.80. The $T_c$ was defined as
the point where the real part deviates from the flattened normal
state part. The narrow transition width is about 1 $\sim$ 2 K for
the annealed samples from overdoped regime to underdoped regime,
which also indicates the high quality of our samples over the whole
superconducting dome.

\begin{figure}
  \center\includegraphics[width=2.8in]{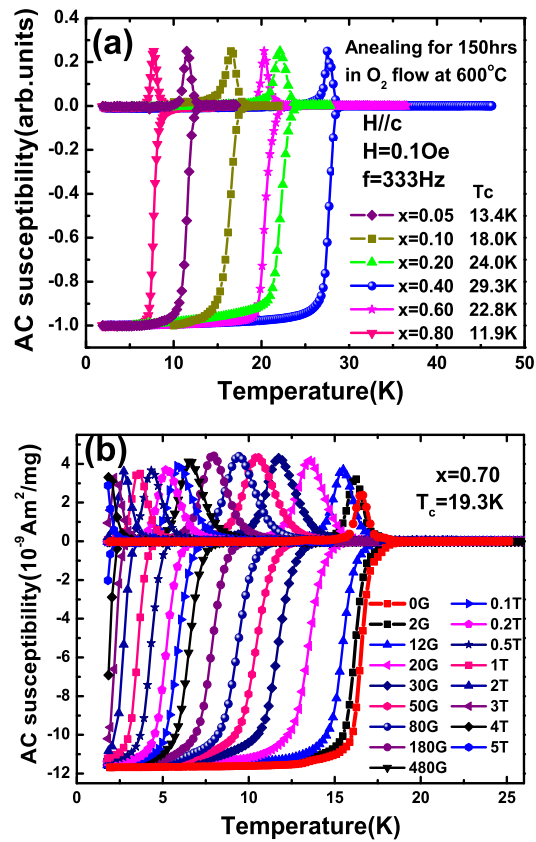}\\
  \caption{(color online) (a). The normalized AC susceptibility for the annealed  single crystals with
  $x$= 0.05, 0.10, 0.20, 0.40, 0.60 and 0.80.
  The $T_c$ was defined as the point where the real part deviates from the flattened normal state part.
  (b). The typical AC susceptibilities under different external DC fields,
  which show the quick suppression of superconductivity under low DC fields.}
  \label{f3}
\end{figure}

According to previous works, the superconducting dome in the phase
diagram of Bi2201 is narrower than the "universal bell shape" shown
as $T_c/T_{cmax}=1-82.6(p-0.16)^2$ \cite{Preslan}.  Thus we deduced
the hole concentration by using the corrected formula in Bi2201,
which was proposed by Ando \etal. in the form of
$T_c/T_{cmax}=1-255(p-0.16)^2$  \cite{Ando3}.  It should be noticed
that the hole concentration $p$  decreases as the La doping level
$x$ increases, thus the substitution of Sr$^{2+}$ with La$^{3+}$
enhances the electron doping and reduces the hole concentration
\cite{Ando2}.  It is found that the hole concentration $p$ for the
samples annealed at 600 $^o$C had a linear relation with La doping
level $x$ [black line in Figure 4(a)], which is consistent with the
previous work \cite{Ando3}.  The hole concentration locates on the
superconducting dome, ranging from heavily overdoped regime to
extremely underdoped side where the superconductivity is absent [red
line in Figure 4(b)]. In addition, the hole concentration shown in
Figure 4 for our Bi-Bi2201 samples is deduced from ARPES
measurements by integrating the whole Fermi surface area after a
tight-binding fitting to the raw data \cite{Pan}.  The data in the
heavily overdoped regime could not be given for the unavailable
samples. For the overdoped sample above $p$=0.18, the nominal
content of Bi for Bi-Bi2201 is less than Sr, which means Bi:Sr$<$1
(or $x <$0). For our experiments, those samples with $x$=0 are very
tiny as needle, and it could not take ARPES measurements on them.
Moreover, it seems impossible to grow sizeable crystals with
Bi:Sr$<$1 by the floating-zone method \cite{Liang1, Liang2, Luo1,
Lin}. Anyway, just like our expectations, the curve of $p$ vs $x$
for Bi-Bi2201 exhibits almost a straight line correlation [blue line
in Figure 4(a)]. However, the slopes of the two lines are different.
Because the strength of A-site disorders is the weakest in La-Bi2201
system and strongest in Bi-Bi2201 system among Bi2201 family, the
$T_c$ is suppressed more quickly by more substitutions on Sr$^{2+}$
sites in the latter. This can be seen clearly in Figure 4(b), where
a distorted dome with a narrower shape shows up in Bi-Bi2201 system
[black line in Figure 4(b)].
\begin{figure}
  \center\includegraphics[width=3.5in]{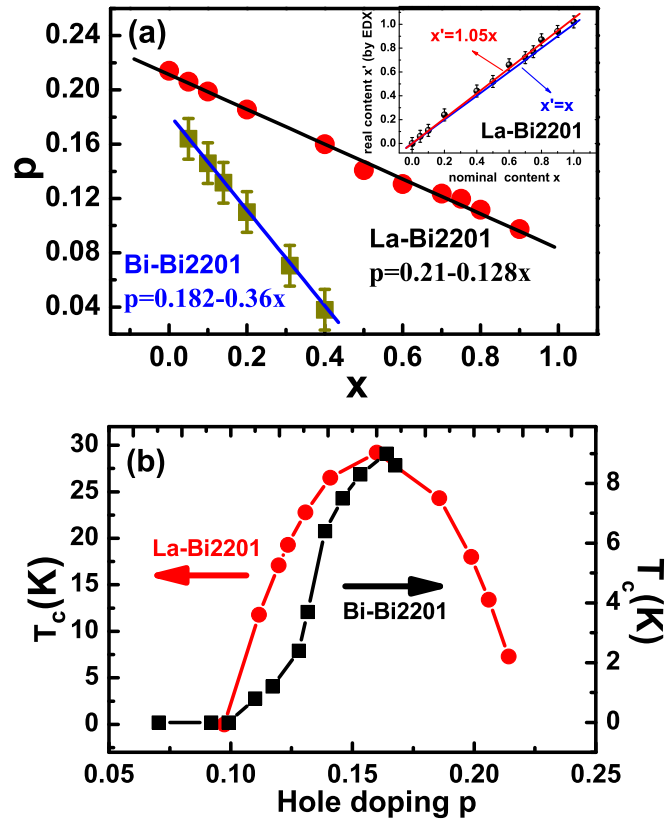}\\
  \caption{(color online)(a). The linear relation between hole concentration $p$ and Ln
  doping level $x$. The solid lines are linear fitting for the scattering points.
  Inset: the real content doping of La for La-Bi2201, which is determined by EDX
  measurements. The crystal composition is very close to the starting
  material, where the segregation coefficient $K=C_s/C_i$ is only about 1.05, slightly above 1.
  (b). The superconducting dome for Bi and La doped Bi2201, where
  the hole concentration is deduced from the linear formulas in Figure 4(a).
  It can be seen that the "bell shape" is distorted in Bi-Bi2201.
  }\label{f4}
\end{figure}
It should be noted that we use the nominal doping content of La (the
value of $x$) in the starting material to represent the composition
of our crystals throughout this paper. Actually, the segregation
always happens for the crystals grown from melt, particularly in
Bi-2201 systems. We then became aware that the content of La in the
single crystals may be quite different from the initial doped
concentration in the polycrystalline powder. So we carried out the
EDX measurements on each crystal with different doping of La, and
obtained the relative percentage for each element in approximate
value. Based on the assumption that the proportion of Sr:La in the
crystals equates to (2-$x'$):$x'$, the real doping level of La $x'$
is deduced and shown in the inset of Figure 4(a). It can be found
that the real content of La $x'$ is slightly larger than the nominal
content $x$ with the relation $x'=1.05x$. Thus the crystal
composition is very close to the starting material, where the
segregation coefficient $K=C_s/C_i$ is only about 1.05, slightly
above 1. So it is reasonable that we use the nominal values of La to
represent the real contents just for convenience.

\subsection{Post-annealing effects on superconductivity}

\begin{figure}
  \center\includegraphics[width=3.0in]{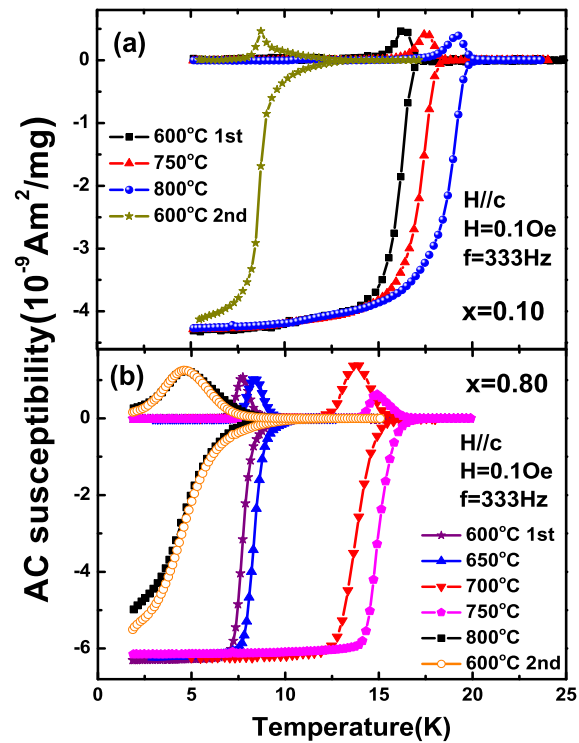}\\
  \caption{(color online) Typical post-annealing effects in the
  sample with $x$= 0.10 (overdoped) and $x$= 0.80 (underdoped) .
   }\label{f5}
\end{figure}
In order to investigate the connection between the superconductivity
and oxygen content, post-annealing experiments were carried out on
our single crystals. For each nominal composition with different La
doping level $x$, a typical as-grown single crystal was selected and
then annealed at different temperatures. All treatments were made
under 1 atm oxygen pressure with flowing rate 30 $\sim$ 50 cc/min
for more than 150 hours.  It was found that the oxygen distribution
in the crystal with moderate size reached the equilibrium state
after annealing for more than 100 hours, because the superconducting
transition and $T_c$ could not be changed any more even annealing
for more time at the same temperature. So 150 hours for each step of
post-annealing are enough. At the end of the treatment, the samples
were quenched to room temperature in air, and then the AC
susceptibility was measured on them. Several post-annealing
temperatures were chosen, changing from 600 $^o$C to 800 $^o$C at 50
$^o$C per step.  As a final check of reversibility, all samples were
annealed at 600 $^o$C again at the final step. We used the same
sample for each nominal doping level of La throughout the whole
procedure.

\begin{figure}
  \center\includegraphics[width=3.0in]{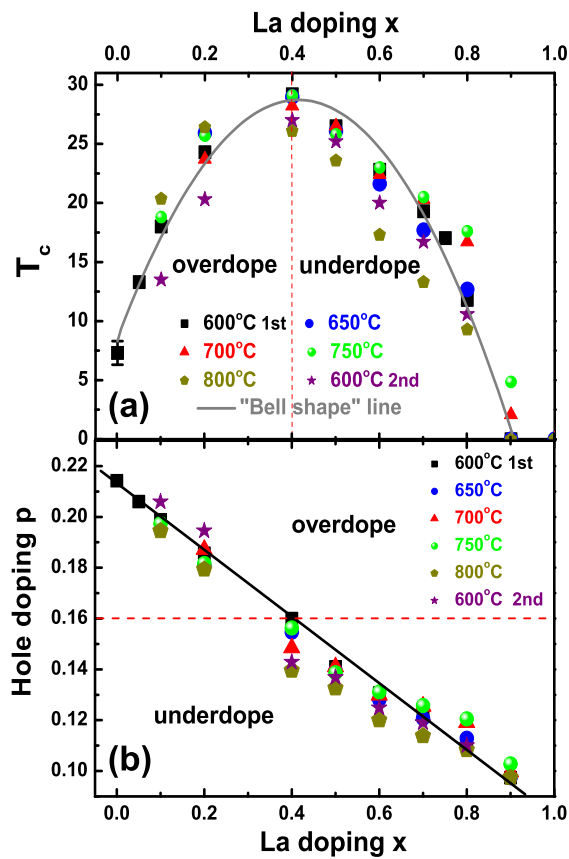}\\
  \caption{(color online) The post-annealing effects on $T_c$ and $p$
  at different temperatures for each La doping level $x$ in La-Bi2201 system.
   }\label{f7}
\end{figure}
Figure 5 shows the post-annealing effects on the samples with $x$=
0.10 and 0.80. Apparently, the superconductivity could be tuned by
post-annealing. For the sample with $x$= 0.10, the $T_c$ increases
as the post-annealing temperature increases.  In the final step of
post-annealing at 600 $^o$C, the superconducting transition could
not return to the shape as that in the first step.  There is a long
tail above the sharp peak of the imaginary part of AC
susceptibility, which indicates that the crystal becomes much
inhomogeneous in the oxygen distribution.  The as-grown crystals
with $x$= 0.80 exhibit no superconductivity above 1.6 K.  However,
after post-annealing at 600 $^o$C for the first step,
superconductivity with $T_c$= 11.9 K emerges, and it can be promoted
by post-annealing at higher temperature 650 $^o$C $\sim$ 750 $^o$C.
The superconducting transition width and the magnitude of
diamagnetic signal only change a little during these steps.
Unfortunately, after post-annealing at 800 $^o$C, $T_c$ drops down
for several Kelvins and the superconducting transition gets much
broader. Although the sample was annealed again at 600 $^o$C at the
last step, the transition and $T_c$ can not be improved much more.
This is similar to the Bi doped Bi2201 system. It seems not easy to
tune the $T_c$ of Bi-Bi2201 by post-annealing
\cite{Vedeneev1,Vedeneev2,Sonder}. According to our previous work,
the Bi-Bi2201 system has no significant post-annealing effects at
any temperatures less than 600 $^o$C, while annealed at higher
temperatures makes the crystals melting down \cite{Luo1}.

The post-annealing effects on $T_c$ are summarized in Figure 6(a)
for each doping in La-Bi2201 system.  Obviously, the magnitude of
$T_c$ varies as the post-annealing temperature changes, and the
detailed behaviors are also various for different La doping levels.
For the optimal doping samples with $x$= 0.40, $T_c$ is very robust
under most post-annealing cases.  However, in the heavily underdoped
regime with $x \geq$ 0.70, $T_c$ is affected easily by
post-annealing. The $T_c$ reaches the maximum magnitude after
post-annealing at 750 $^o$C in the underdoped regime. While in the
overdoped regime, the highest $T_c$ could be obtained by
post-annealing at 800 $^o$C. Unfortunately, the result of
post-annealing at 800 $^o$C is anomalous for all underdoped samples,
which gives a much lower $T_c$ and a broader transition. The "bell
shape" is distorted in the case of post-annealing at 800 $^o$C, too.
Moreover, after post-annealing at 600 $^o$C again at the final step,
the $T_c$ can not return to the magnitude at the first step. This
suggests that the crystal structure may distort at such high
temperatures being close to its melting point.  It is more serious
in the overdoped regime, since the melting points for these samples
are lower than those in the underdoped side.  The $T_c$ after
post-annealing at 600 $^o$C for the final step drops to much lower
level, and the long-tail feature in AC susceptibility is also very
clear. Furthermore, if we plot the hole concentration $p$ for each
$T_c$ in the same La doping level, it could be found that the hole
concentration related to the oxygen content also varies with
different post-annealing temperature [Figure 6(b)], where $p$ is
determined by the formula from Ando \etal. \cite{Ando3} using
$T_{cmax}= 29.2$ K.  For the first step of post-annealing at 600
$^o$C, the relation between $p$ and $x$ is almost linear.  It is
also deviated seriously in the case of post-annealing at 800 $^o$C.
After the second step of post-annealing at 600 $^o$C, the hole
concentration $p$ could not return to the same level as the first
step of post-annealing at 600 $^o$C. It increases in the overdoped
regime, while the case is opposite in the underdoped regime. These
different effects of post-annealing suggest that the nature of
superconducting mechanism in the overdoped regime may not be the
same as the underdoped regime \cite{Lee}. One typical example is
that the phase separation may exist in the overdoped regime
\cite{Uemura}, where it is absent in the underdoped regime
\cite{Wen1,Wen2}.

\section{Summary}
In summary, we have successfully grown high-quality
Bi$_2$Sr$_{2-x}$La$_x$CuO$_{6+\delta}$ (0$\leq$x$\leq$1.00) single
crystals by TSFZ method.  The sharp peaks in the patterns of XRD
indicate high crystalline quality for our samples. After
post-annealing in the flowing oxygen at 600 $^o$C, The results of AC
susceptibility show sharp superconducting transitions. The hole
concentration $p$ exhibits a linear relation with La doping level
$x$. It ranges from the heavily overdoped regime ($p \approx$ 0.2)
to the extremely underdoped side ($p \approx$ 0.08) where the
superconductivity is absent. Comparing between the La-Bi2201 and
Bi-Bi2201 system, the superconducting dome has a narrower shape in
the latter due to the stronger effects from A-site disorders. The
hole concentration could also be tuned by post-annealing at
different temperatures in flowing oxygen. The effects of
post-annealing on $T_c$ and $p$ are summarized over the whole
superconducting regime of the phase diagram, which indicate the
nature of superconductivity in overdoped regime may be different
from that in underdoped regime.

\textbf{Acknowledgements}

This work was financially supported by the Natural Science
Foundation of China, the Ministry of Science and Technology of China
(973 Projects Nos. 2006CB601000, 2006CB921802 and 2006CB921300), and
Chinese Academy of Sciences (Project ITSNEM).  The authors
acknowledge the ARPES data measured by Mr. Zhihui Pan and Prof. Hong
Ding in Boston College and IOP, CAS, and the helpful discussions
with Prof. Lei Shan and Prof. Cong Ren at IOP, CAS.


\section*{References}
\begin{thebibliography}{30}

\bibitem{Lee} Lee P A, Nagaosa N and Wen X G 2006 \RMP \textbf{78} 17

\bibitem{Preslan} Presland M R, Tallon J L, Buckley R, Liu R S and Flower N 1991 {\it Physcica} C \textbf{176} 95

\bibitem{Eisaki} Eisaki H, Kaneko N, Feng D L, Damascelli A, Mang P K, Shen K M, Shen Z X and Greven
M 2004 \PR{B} \textbf{69} 064512

\bibitem{Fujita} Fujita K, Noda T, Kojima K M, Eisaki H and Uchida S 2005 \PRL \textbf{95} 097006

\bibitem{Okada} Okada Y, Takeuchi T, Baba T, Shin S, Ikuta H 2008 {\it J. Phys. Soc. Jpn.} \textbf{77} 074714

\bibitem{Slezak} Slezak J A \etal 2008 {\it PNAS} \textbf{105} 3203

\bibitem{Michel} Michel C, Hervieu M, Borel M M, Grandin A, Deslandes F, Provost J and Raveau B 1987 \ZP{B}
\textbf{68} 421

\bibitem{Fleming} Fleming R M, Sunshine S A, Schneemeyer L F, Van Dover R B,
Cava R J, Marsh P M, Waszczak J V, Glarum S H, Zahurak S M and
DiSalvo F J 1990 {\it Physcica} C \textbf{173} 37

\bibitem{Martovitsky} Martovitsky V P, Gorina J I, and Kaljushnaia G A
1995 \SSC \textbf{96} 893

\bibitem{Gorina} Gorina Y I, Kalyuzhnaya G A, Rodin V V, Sentyurina N N,
Stepanov V A and Chernook S G 2007 {\it Crystallography Reports}
\textbf{52(4)} 735

\bibitem{Liang1} Liang B, Maljuk A and Lin C T 2001 {\it Physcica} C
\textbf{361} 156

\bibitem{Liang2} Liang B and Lin C T 2004 {\it J. Crystal Growth}
\textbf{267} 510

\bibitem{Ando1} Ando Y and Murayama T 1999 \PR{B} \textbf{60} R6991

\bibitem{Luo1} Luo H Q, Fang L, Mu G and Wen H H 2007 {\it J. Crystal Growth} \textbf{305} 222

\bibitem{Lin} Lin C T, Freiberg M, Sch$\ddot{o}$nherr E 2000 {\it Physcica} C
\textbf{337} 270

\bibitem{Ando2} Ono S and Ando Y 2003 \PR{B} \textbf{67} 104512

\bibitem{Ando3} Ando Y, Hanaki Y, Ono S, Murayama T, Segawa K, Miyamoto N and Komiya S
2000 \PR{B} \textbf{61} R14956

\bibitem{Pan} Pan Z H \etal pre-print on Arxiv: 0806.1177

\bibitem{Vedeneev1} Vedeneev S I, Jansen A G M, Haanappel E and Wyder P 1999 \PR{B} \textbf{60} 12467

\bibitem{Vedeneev2} Vedeneev S I and Maude D K 2004 \PR{B} \textbf{70} 184524

\bibitem{Sonder} Sonder F, Chakoumakos B and Sales B 1989 \PR{B} \textbf{40} 6872

\bibitem{Uemura} Uemura Y J \emph{et al.} 1993 {\it Nature} \textbf{364} 605

\bibitem{Wen1} Wen H H, Yang W L, Zhao Z X and Ni Y M 1999 \PRL \textbf{82} 410

\bibitem{Wen2} Wen H H 2000 {\it PNAS} \textbf{97} 11145

\end{thebibliography}
\end{document}